\documentstyle[12pt]{article}

\newcommand{\be}{\begin{equation}}
\newcommand{\ee}{\end{equation}}
\newcommand{\bea}{\begin{eqnarray}}
\newcommand{\eea}{\end{eqnarray}}

\begin{document}
\begin{center}

{\Large\bf Analytic Approach to  Perturbative  QCD}

 \medskip

{\bf B.A. Magradze$^\star$
\footnotetext {$^\star$E-mail address:
magr@rmi.acnet.ge}}
\medskip

{\it A. Razmadze  Mathematical Institute,
Georgian Academy of Sciences,\\
Tbilisi 380093
Georgia\\}

\medskip

{\footnotesize \bf Abstract}
\medskip

\parbox{110mm}{\footnotesize
The two-loop invariant (running) coupling  of QCD is written
in terms of the Lambert W function.
The analyticity structure
of the  coupling in the complex $Q^2$-plane is established.
The corresponding analytic coupling is
reconstructed via a dispersion relation.
We also consider some other approximations to the QCD $\beta$-function,
when the corresponding couplings are solved in terms of the
Lambert function.
The Landau gauge gluon propagator
has been considered in the renormalization group
invariant analytic approach (IAA).
It is shown that  there  is a nonperturbative ambiguity
in determination of the anomalous dimension function of the gluon field.
Several analytic solutions for the propagator at the one-loop order
are constructed. Properties of the obtained analytical
solutions  are
discussed.}
\end{center}

 \medskip

\section{Introduction}
One of the most important tasks in  quantum chromodynamics (QCD)
is to find the momentum dependence
of the invariant (running) coupling   $ {\bar{\alpha}_{s}}(Q^2) $.
For large $Q^2$, the perturbative approximations of the  coupling
are reliable, since the theory is asymptotically free.
Outside of the asymptotic region, the perturbative invariant
coupling is in fact large.
Moreover, it has wrong analytical properties:
the unphysical Landau singularities \cite{red,bog}
appear at small space-like momenta.
As a consequence,
renormalization group (RG) improved
expressions for
perturbation theory (PT) approximations of physical quantities
also have incorrect analytical properties.
In particular, the electroweak current-current correlation functions
parametrized by the running coupling  in  PT
do not obey  dispersion relations (DRs).

 DRs are fundamental for proving many important results
in quantum field theories \cite{bsh, btv, o}.
There exists a vast theoretical literature for the application of
DRs in deep-inelastic lepton-hadron scattering. This problem has
been studied in rigorous manner on the basis of  Jost-Lehmann-Dyson
representation \cite{btv}.
Other notable examples are
the $e^{+}e^{-}$ annihilation ratio $R_{e^{+}e^{-}}$, and the ratio
of hadron to leptonic $\tau$-decay widths $ R_{\tau} $.
These time-like quantities can be related to the hadronic two-point
correlation functions through DRs.
These  DRs  provide a well-defined method for definition of
the running coupling  for time-like momentum.

In the infrared region (IR), the investigation   encounters great
difficulties due to a failure of  PT.
Several methods developed based on  dispersion relation \cite{dmw, gr},
Borel summation and renormalons \cite{gr,nb},
renormalization scheme (RS) choices \cite{kp} and
the Schwinger-Dyson equations (SDEs) \cite{dl}.
Common belief is that the unphysical singularities could be eliminated
when one adds higher order perturbative and nonperturbative corrections
\cite{bsh,bsh1}.

The widespread supposition is  that infrared properties of the
exact effective coupling may lead to color confinement in QCD.
Most of the suggestions of a confinement mechanism imply the
assumption that the invariant coupling  has a
singularity at $Q^2=0$. For instance, in the linear confinement picture,
it is assumed that $\bar \alpha_{s}(Q^2) \sim Q^{-2}$, $Q^2\rightarrow 0$.
So   the Landau pole moves out of
the space-like region  and  causality is restored \cite{mn}.
It was confirmed that this  asymptotic behavior is consistent with
the  SDEs and Slavnov-Taylor identities of QCD \cite{mn,bbz,ar}.

On the contrary,
other authors have argued  that the invariant coupling
should be finite in the IR region.
It is assumed  that the coupling ``freezes'' at low energies due to
an infrared-stable fixed point of the theory.
Such a possibility has been occurred in the K scheme \cite{gr1,kkp},
where the corresponding $\beta$-function in the three-loop approximation
acquires an infrared fixed point.
The similar conclusion has been achieved using  Pade approximant methods
\cite{egks} or the Banks-Zaks expansion \cite{st}.
Note that infrared ``freezing''can not be in accord with the above mention
linear confinement mechanism.  However, there is   an approach to
confinement based on the BRST algebra and the RG methods
(the so-called metric confinement)
\cite{o1,nis}.
A linear quark-antiquark potential is not implied
within this framework.
Instead,  an approximately linear potential is supposed
\cite{nio}.

One possible resolution of the ``Landau-ghost'' problem  has been
suggested long time ago in Ref. \cite{red}.
In Ref.\cite{bog}, the idea to combine the RG invariance and
the $Q^2$-analyticity (in the context of quantum
electrodynamics (QED)) have been proposed.
A RG invariant analytic version of PT
has been elaborated in this work.
Recently, this method has been  successfully applied in QCD.
It was shown, that the analytic running coupling is stable for the
whole interval of momentum, and has an universal infrared limit at
$Q^2=0$ \cite{ss}.
The ``analyticized'' perturbative approximations, as
opposed to standard PT, exhibit reduced RS dependence
and the results are not sensitive to higher-loop corrections
\cite{rs}.
These features of  RG invariant analytic approach (IAA)
have been
demonstrated in recent calculations
of $R_{e^{+}e^{-}}$ \cite{epl}
and $ R_{\tau} $ \cite{tau}.
The calculations have been
performed to next-to-next-to-leading order.
Other improvement of the analytic approach is that it provides
a self-consistent determination of the running coupling
in the time-like region \cite{mink}.
In Refs.~\cite{bjo} the Bjorken  and the Gross-Llewellyn
Smith sum rules in IAA are considered.
The Landau gauge gluon propagator in IAA
has been studied in Ref.\cite{my}.
For most recent applications of the analytic approach to QCD see
the recent review \cite{ss1}.

In Sec.2 we consider  some technical aspects of the RG IAA.
The two-loop running coupling is solved explicitly in terms of the
Lambert W function. This allows us to investigate analytic structure of
the coupling in the complex $Q^2$ plane.
The corresponding analytic
solution is reconstructed via the dispersion relation.
To discuss the RS dependence,
we compare the exact expressions for the causal coupling
in the two different renormalization schemes.
In Sec.3 the structure of the $\beta$-function in APT is studied and
other  approaches to the Landau pole problem  are discussed.
The special (nonperturbative) model $\beta$-function is investigated.
In this case, we also obtain the explicit solution for the coupling
in terms of the Lambert W function and show that the solution
is consistent with causality.
In Sec.4
we apply the analytic approach to
the gluon propagator in the Landau gauge.
Several ways of restoring analyticity for the propagator are considered.
The ambiguity in causal ``analyticization procedure''
is discussed \cite{sh1}. Sec.5 contains the concluding remarks.

\section{Lambert's W function in IAA}
The running coupling  of massless QCD satisfies
the differential equation \footnote{We use the notation $Q^2=-q^2$, $Q^2>0$
corresponds to a spacelike momentum transfer.}
\be
Q^2\frac{d}{dQ^2}{\bar{\alpha}_{s} }(Q^2)=
\beta({\bar{\alpha}_{s} }(Q^2)),
\label{eq:eff}
\ee
with the initial condition
\be
{\bar{\alpha}_{s} }({\mu}^2)={\alpha}_{s},
\label{nrm}
\ee
here ${\alpha}_{s}=\frac{g^2}{4\pi}$, $g$ is the renormalized
coupling constant, and $\mu$ is the renormalization point.
We assume that
the non-trivial $\beta$ function exists with the perturbative
asymptotic expansion in powers of ${\alpha}_{s}$
\be
\label{beta1}
\beta({\alpha}_{s})\approx-\frac{\beta_0}{4\pi}({\alpha}_{s})^2
-\frac{\beta_1}{(4\pi)^2}({\alpha}_{s})^3
-\cdots,
\ee
the first two coefficients of the formal power series are independent
of the chosen renormalization conditions.
Their values are
\be
\label{coef}
\beta_{0}=\left(11-\frac{2}{3}N_{f}\right),\hspace{5mm}
\beta_{1}=\left(102-\frac{38}{3}N_{f}\right),
\ee
with $N_{f}$ being  the number of quark flavors.
The solution of  Eq.~(\ref{eq:eff}), which satisfies the initial condition
(\ref{nrm}), has the form \cite{stev1}
\be
\label{pt}
\ln\left(\frac{Q^{2}}{{\Lambda}^2}\right)=
\frac{4\pi}{\beta_{0}{\bar\alpha}_{s}(Q^2)}-
\frac{\beta_1}{\beta_0^2}\ln\left(1+
\frac{4\pi\beta_0}{\beta_1{\bar{\alpha}_{s}(Q^2)}}\right)
+\psi(\bar{\alpha}_{s}(Q^2)),
\ee
where
$$
\psi({\alpha}_{s})= \int_{0}^{{\alpha}_{s}}\left(\frac{1}{\beta(x)}
-\frac{1}{\beta^{(2)}(x)}\right)dx,
$$
and ${\beta}^{(2)}(x)$ denotes the two-loop $\beta$-function.
The QCD scale parameter $\Lambda$ here is different from the
conventional one ${\Lambda}_{\overline{MS}}$, so that
$\Lambda=(b)^{(\frac{-b}{2})}{\Lambda}_{\overline{MS}}$ with
$b=\frac{\beta_{1}}{\beta_{0}^2}$ \cite{stev1}.
To obtain explicit expressions for $\bar\alpha_{s}(Q^2)$
as a function of $Q^2$,
beyond the one-loop order, one must solve the transcendental
equation (\ref{pt}).

Let $\bar\alpha^{(n)}(Q^2)$ be  nth-order perturbative
approximation to $\bar\alpha_{s}(Q^2)$ determined implicitly by Eq.~(\ref{pt}).
Then the corresponding analytic running coupling  can be defined via the
K\"{a}llen-Lehmann
integral \cite{bog,ss}
\be
\label{kl}
\bar\alpha_{an}^{(n)}(Q^2) =\frac{1}{\pi}\int_{0}^{\infty}
\frac{{\rho}^{(n)}(\sigma)}{(\sigma+Q^2-i0)}d\sigma,
\ee
the spectral function $ {\rho}^{(n)}(\sigma) $ is the discontinuity
of the ``initial'' expression
$ \bar\alpha^{(n)}(Q^2) $ along the negative $Q^2$ axis:
${\rho}^{(n)}(\sigma)=Im \bar\alpha^{(n)}(-\sigma-i0) $.
The remarkable result is that the limiting value  $\bar\alpha_{an}(0)$
is universal
and determined only by one-loop calculation \cite{ss}:
$$
\bar\alpha_{an}^{(n)}(0)=\bar\alpha_{an}^{(1)}(0)=\frac{4\pi}{\beta_{0}}.
$$
By using Cauchy's theorem Eq.~(\ref{kl}) can be rewritten as follows
\be
\label{anc2}
\bar\alpha_{an}^{(n)}(Q^2)={\bar\alpha}^{(n)}(Q^2)+
{\theta}^{(n)}(Q^2).
\ee
Here the ``nonperturbative''
term ${\theta}^{(n)}$
(which compensates the unphysical
contributions
of  $\bar\alpha^{(n)}$)
comes from the ghost cut
\be
\label{theta}
{\theta}^{(n)}(Q^2)=
\frac{1}{\pi}\int_{k_{L}^2}^{0}
\frac{{\rho}_{\theta}^{(n)}(\sigma)}{(\sigma+Q^2-i0)}d\sigma,
\ee
where $k_{L}^2<0$, to first and second orders $k_{L}^2=-{{\Lambda}}^2$.

Let us now consider the two-loop running coupling in more
detail. It is more convenient to introduce  the quantity
\be
\label{newdef}
\begin{array}{ccc}
a(x)=\frac{\beta_{0}}{4\pi}\bar\alpha(Q^2),&
$where$ & x=\frac{Q^2}{{\Lambda}^2}.\\
\end{array}
\ee
To second order Eq.~(\ref{pt})  reads
\be
\label{trans}
\frac{1}{a^{(2)}(x)}-b \ln\left(1+\frac{1}{b}\frac{1}{a^{(2)}(x)}
\right)=\ln x,
\ee
where we have denoted
$b=\frac{\beta_1}{{\beta_0}^2}$
$(b=\frac{64}{81}$ for $ N_{f}=3)$. In what follows we shall consider
the phenomenologically interesting case when
$N_{f}\le 8.05$ ($0 \le b <1$).  Then    the
two-loop running coupling has the ghost singularity on the positive $x$-axis.
The transcendental equation (\ref{trans}) can be solved   by the
iteration method.
The solution given by one iteration is
\footnote {Historically, for the first time the 2-loop iterative solution
has been proposed
in Ref.~\cite{bsh3} in the context of QED.}
\be
\label{iter}
a^{(2)}_{it}(x)=\frac{1}{\ln x+b \ln\left(1+\frac{1}{b}\ln x\right)}.
\ee
Consider function (\ref{iter}) in the whole
complex $x$-plane. It has the Landau pole at $x=1$, the logarithmic
branch point at $x=exp(-b)$ and standard branch point at $x=0$.
The corresponding analytic coupling is given by the K\"{a}llen-Lehmann
integral (see Refs.\cite{ss})
\be
\label{kl1}
a^{(2)}_{it.an}(x)=\frac{1}{\pi}\int_{0}^{\infty}\frac{\bar
\rho_{it}^{(2)}(s)}{s+x}ds
=\frac{1}{\pi}\int_{-\infty}^{\infty}\frac{e^{t}}{e^{t}+x}\tilde
\rho_{it}^{(2)}(t) dt,
\ee
where $\tilde\rho^{(2)}_{it}(t)\equiv\bar\rho^{(2)}_{it}(e^t)\equiv
\rho^{(2)}_{it}(e^{t}{\Lambda}^{2})=
\frac{L_{1}}{(L_{1}^{2}+L_{2}^{2})}$ with
$$
L_{1}=\pi+b \arccos(\frac{b+t}{r}),\hspace{0.3cm}
L_2=t+b \ln(\frac{r}{b}),\hspace{0.3cm} r=\sqrt{(b+t)^2+\pi^2}.
$$
One can investigate  Eq.(\ref{trans}) using the lagrange inversion formula.
Then one finds  that the  solution has second order branch point at
$x=1$: $a^{(2)}(x)\sim \frac{1}{\sqrt{2b(x-1)}}$.
Hence,  approximation (\ref{iter}) violates the analytical properties
of $a^{(2)}(x)$ near
the point
x=1.
In fact, the transcendental Eq.~(\ref{trans}) is exactly solvable
\cite{my1, my2, ggk}.  The solution is
\bea
\label{w}
a^{(2)}(x)=-\frac{1}{b}\frac{1}{1+ \omega(x)}:&
\omega(x)=W(\zeta),
\eea
where
\be
\label{zeta}
\zeta=-\frac{1}{e}x^{-\frac{1}{b}}=
exp\left(-\frac{\ln(x)}{b}-1+i\pi\right),
\ee
and $W(\zeta)$ denotes the Lambert W function \cite{lamb}. This is
the multivalued inverse of
$$
\zeta=W(\zeta)\exp W(\zeta).
$$
The branches of W are denoted $W_{k}(\zeta), k=0,\pm 1,\ldots .$
A detailed review of  properties and
applications of this special  function can be found in
\cite{lamb}.
We shall define branches of W following  this work.
Then the branch cuts are chosen as $\{\zeta:-\infty<\zeta\leq-\frac{1}{e}\}$
and
$\{\zeta:-\infty<\zeta\leq 0 \}$ \footnote{These notations and
definitions of branches of W are different from that used in Refs.\cite{my1}}.
The branch $W_{-1}(\zeta)$ satisfies
$W_{-1}(\zeta)\le-1$ for $-e^{-1}\leq \zeta \leq 0 $.
Now our task is to choose the suitable branches of $W$ in
Eq.~(\ref{w}).
The solution $a^{(2)}(x)$ is supposed to have the following properties:\\
i. It would  be an analytic function in the cut $x$-plane with cuts
$\{x:-\infty<x<0\}$ and $\{x:-\infty< x \le 1\}$.\\
ii. It is real and positive for a real positive $x>1$.\\
iii. $a^{(2)}(x)\sim 1/\ln x $ if $x\rightarrow\infty$ along all
directions in the complex x-plane.\\
It is obvious, that for real positive $ x>1$ the relevant branch is
$\omega(x)=W_{-1}(\zeta)$.
We look for the analytical continuation of $\omega(x)$ to the whole complex
$x$ plane.
To perform the analytical continuation, one has to take into
account the ranges of the branches of
$w=W(\zeta)$ \cite{lamb}.
Let us consider the cases  $0<N_{f}<6.19$ $(0.5\le b<1)$ and
$6.19\le N_{f}\le 8$ $(0<b<0.5)$  separately:

1.  $0\leq N_{f}<6.19$. Evidently, on the upper half-plane $(Im(x)>0)$ the
solution is determined by  $W_{-1}(\zeta)$.
On the lower half-plane the branch should be changed.
Indeed, we have crossed the cut in the complex $\zeta$-plane.
Accordingly,  the replacement
$\arg(\zeta)\Rightarrow \arg(\zeta)-2\pi$ is implied.
We conclude that  here the relevant branch is $W_{1}(\zeta)$.
With this choice, the solution is continuous on the line
$1<x<\infty$. Thus we obtain
\be
\omega(x)=\left\{\begin{array}{lll}
W_{-1}(|\zeta|e^{i\varphi_{1}}):
& \varphi_{1}=\pi-\frac{1}{b}\arg(x) & \mbox{if $ 0<\arg(x)\le\pi.$} \\
W_{1}(|\zeta|e^{i\varphi_{2}}):
& \varphi_{2}=-\pi-\frac{1}{b}\arg(x) & \mbox{if $ -\pi<\arg(x)\le 0.$}
\end{array}
\right.
\label{eq:w1}
\ee
Here $ |{\zeta}|={\frac{1}{e}}|x|^{-\frac{1}{b}} $.

2. $6.19\leq N_{f}\leq 8$. In this case, we separate the
upper half-plane, $(\arg(x)\in(0,\pi))$, into sectors
$$
\begin{array}{lll}
2(n-1)b\pi<\arg(x)\leq \min(2nb\pi,\pi),&
1\leq n \leq n_{max},&  \frac{1}{2b}\leq n_{max}<\frac{1}{2b}+1, \\
\end{array}
$$
(for $N_{f}=8$,  $n_{max}=25$).
An image of a sector is a sheet of the Riemannian
surface of $\zeta$, which maps onto the range of a branch of
$\omega(x)=W(\zeta)$.
The analytical continuation to the n-th sector
can be written as
\be
\begin{array}{cc}
\omega(x)=W_{-n}(|\zeta|e^{i\varphi_{-n}}),
& \mbox{where $\varphi_{-n}=(2n-1)\pi-\frac{1}{b}\arg(x)$}.\\
\end{array}
\ee
To obtain the solution on the lower half-plane,
($\arg(x)\in(-\pi,0)$), one must use the
formula $W_{n}(\overline{\zeta})=\overline{W_{-n}(\zeta)}$ \cite{lamb}.

Now we can  reconstruct the corresponding analytic coupling
$a_{an}^{(2)}(x)$.
\begin{table}[h]
\caption{The $Q^2$ dependence of the analytic running couplings
$  \alpha_{an}^{(2)}(Q^2)$, $  \alpha_{it.an}^{(2)}(Q^2)$
and $\alpha_{an}^{spec}(Q^2)$ for $N_{f}=3$.
Here, we have used the reference value of $\bar\alpha_{s}({M_{\tau}}^2)=0.36$
for $M_{\tau}=1.777 GeV$.}
\vspace{0.2cm}
\begin{center}
\footnotesize
\begin{tabular}{|l|l|l|l|c|l|l|l|l|}
\hline
$ Q^2$ $ GeV^2 $   &
$  \alpha_{an}^{(2)}(Q^2)$ &
$  \alpha_{it.an}^{(2)}(Q^2)$ &
$\alpha_{an}^{spec}(Q^2)$  &
                        &
$ Q^2$ $ GeV^2 $ &
$  \alpha_{an}^{(2)}(Q^2)$ &
$  \alpha_{it.an}^{(2)}(Q^2) $ &
$\alpha_{an}^{spec}(Q^2) $ \\
\hline
0     & 1.396    & 1.396  & 1.396 & & 5   & 0.330 & 0.330   & 0.330  \\
0.001 & 0.985    & 0.969  & 0.980 & & 6   & 0.319 & 0.319   & 0.319  \\
0.01  & 0.846    & 0.831  & 0.838 & & 7   & 0.310 & 0.310   & 0.310  \\
0.05  & 0.716    & 0.706  & 0.710 & & 8   & 0.302 & 0.302   & 0.303  \\
0.1   & 0.654    & 0.647  & 0.649 & & 9   & 0.296 & 0.296   & 0.296  \\
0.5   & 0.507    & 0.505  & 0.505 & & 10  & 0.290 & 0.290   & 0.291   \\
1     & 0.447    & 0.446  & 0.446 & & 20  & 0.256 & 0.257   & 0.257  \\
2     & 0.393    & 0.392  & 0.392 & & 30  & 0.239 & 0.239   & 0.240  \\
3     & 0.364    & 0.364  & 0.364 & & 40  & 0.228 & 0.228   & 0.229  \\
4     & 0.344    & 0.344  & 0.344 & & 50  & 0.220 & 0.220   & 0.221  \\
\hline
\end{tabular}
\end{center}
\end{table}
In the case $0\leq N_{f}\leq 6$,
with Eqs.~(\ref{kl}), (\ref{w}) and (\ref{eq:w1}), we obtain
\be
\label{kl2}
a^{(2)}_{an}(x)=\frac{1}{\pi}\int_{0}^{\infty}\frac{\bar\rho^{(2)}(s)}{s+x}
ds=
\frac{1}{\pi}\int_{-\infty}^{\infty}\frac{e^{t}}{(e^{t}+x)}{\tilde
\rho^{(2)}(t)}dt,
\ee
where $\tilde\rho^{(2)}(t)\equiv\bar\rho^{(2)}(e^t)
\equiv\rho^{(2)}(e^{t}
\Lambda^{2})$
and
\be
\begin{array}{cc}
\tilde\rho^{(2)}(t)=-\frac{1}{b}Im\left(\frac{1}{1+
W_{1}(\zeta_{1}(t))}\right),
&\zeta_{1}(t)=\exp\left(-\frac{t}{b}-1+i(\frac{1}
{b}-1)\pi\right).
\end{array}
\label{eq:ro}
\ee
Equivalently, using the Cauchy formula we can rewrite Eq.~(\ref{kl2}),
in the form
$$
a^{(2)}_{an}(x)=a^{(2)}(x)-
\frac{1}{\pi}\int_{0}^{1}\frac{1}{s-x}Im\{a^{(2)}(s+i0)\}ds.
$$
In particular, for $0\leq x \leq 1$, we have
$$
a^{(2)}_{an}(x)=Re\{a^{(2)}(x+i0)\}-
PV\frac{1}{\pi}\int_{0}^{1} \frac{1}{s-x}Im\{a^{(2)}(s+i0)\}ds,
$$
where PV denotes the principal value of the integral, and
$$
a^{(2)}(x+i0)=-\frac{1}{b}\frac{1}{1+
W_{-1}\left(-\frac{1}{e}x^{-\frac{1}{b}}\right)}.
$$
Numerical results for the exact two-loop coupling (\ref{kl2}) as well
as for the iterative solution (\ref{kl1}) are summarized in the Table 1.
The relative error for the iterative solution (\ref{kl1}),
is less then 1.8 \% for the considered
interval. This gives 8\% error in the value of the QCD scale
parameter
${\Lambda}^{(2)}$ for $N_{f}=3$ (see Table 2). Following
Refs.\cite{ss} we may use the average
$$
A(Q)=\frac{1}{Q}\int_{0}^{Q}\bar{\alpha}_{s}(\mu^2)d\mu
$$
for comparison. For $ Q=2 $ GeV,  we find that iterative solution
(\ref{kl1})
gives answer for $A(Q)$ to much better than $ 0.5\% $ accuracy. The reference
values of
$ \alpha_{an}({M_{\tau}}^2) $ are taken as:
$ \alpha_{an}({M_{\tau}}^2)=0.36 \pm 0.02 $ with  $ M_{\tau}=1.777$ GeV.
\bigskip
\begin{table}
\begin{center}
\caption{Numerical results for $\Lambda$ and $ A(2GeV)$
with exact and iterative solutions.}
\vspace{0.2cm}
\begin{tabular}{|l|l|l|l|}\hline
$ \bar\alpha_{s}({M_{\tau}}^2) $& 0.34 & 0.36 &0.38 \\ \hline
${\Lambda}^{(2)}_{it}$(GeV) & 0.606 &0.706 & 0.815\\
${\Lambda}^{(2)}$(GeV) & 0.665 & 0.772 & 0.889\\
${\Lambda}_{spec}$(GeV)& 0.644 & 0.750 & 0.865\\
$A_{it}^{(2)}(2GeV)$& 0.476 & 0.499 & 0.522\\
$A^{(2)}(2GeV)$& 0.479  & 0.502 & 0.525 \\
$A_{spec}(2GeV)$& 0.477  & 0.500 & 0.523 \\
\hline
\end{tabular}
\end{center}
\end{table}
The instructive example is given by the $\beta$-function of the special
RS
\be
\label{lsb}
\beta(\alpha_{s})=-\frac{\beta_{0}}{4\pi}\frac{{\alpha_{s}}^2}
{1-\frac{\beta_{1}}{4\pi\beta_{0}}\alpha_{s}},
\ee
in this scheme the inverse $\beta$-function contains only two terms
(for application of this RS see.\cite{byz}). Note that expression
(\ref{lsb}) is beyond the formal framework of PT where only
finite order polynomials in the coupling are allowed \cite{kp}.
The RG    equation (\ref{eq:eff}) with (\ref{lsb}) yields
the implicit solution for the coupling
\be
\begin{array}{ccc}
\label{impl}
x=(ba(x))^{b}\exp(\frac{1}{a(x)}):& x=\frac{Q^2}{{\Lambda}^2},&
b=\frac{\beta_{1}}{{\beta_{0}}^2},
\end{array}
\ee
here $\Lambda$ we define following Ref.\cite{stev1}. An inversion of
(\ref{impl}), can be written in terms of the Lambert W function
\be
\begin{array}{ccc}
\label{expl}
a(x)=-\frac{1}{b\omega_{1}(x)}:& \omega_{1}(x)=W(z),&z=-x^{-\frac{1}{b}}.
\end{array}
\ee
For $x>e^{b}$, ($b>0$, for $N_{f}\le 8$) the physical branch is
$W_{-1}(z)$. This branch is  real and yields the correct ultraviolet
behavior for the $a(x)$. We see that $a(x)$ has the branch point
(the unphysical singularity) at $x=e^{b}$, and   is finite at this
point
$$
\begin{array}{cc}
a(x)\sim\frac{1}{b}-\frac{1}{b}\sqrt{\frac{2(x-e^{b})}{be^{b}}},
&x\sim e^{b}.
\end{array}
$$
Let us now perform the analytic continuation in the complex $x$-plane.
For $0\le N_{f}\le 6 $ we obtain
\be
\omega_{1}(x)=\left\{\begin{array}{lll}
W_{-1}(|z|e^{i\varphi_{+}}):
& \varphi_{+}=\pi-\frac{1}{b}\arg(x) & \mbox{if $ 0<\arg(x)\le\pi.$} \\
W_{1}(|z|e^{i\varphi_{-}}):
& \varphi_{-}=-\pi-\frac{1}{b}\arg(x) & \mbox{if $ -\pi<\arg(x)\le 0.$}
\end{array}
\right.
\label{cnt}
\ee
The corresponding causal coupling is defined via the DR
\be
\label{kl3}
{\tilde {a}}_{an}(x)=\frac{1}{\pi}\int_{0}^{\infty}\frac{{\bar\rho}_{1}(s)}{s+x}
ds=
\frac{1}{\pi}\int_{-\infty}^{\infty}\frac{e^{t}}{(e^{t}+x)}
{\tilde\rho}_{1}(t)dt,
\ee
where
$$
{\tilde\rho}_{1}(t)={\bar\rho}_{1}(e^{t})=
-\frac{1}{b}Im\left(\frac{1}
{W_{1}\left(e^{-\frac{t}{b}+i\pi(\frac{1}{b}-1)}\right)}\right)
$$
Let us compare the exact results (\ref{kl2}) and (\ref{kl3})
(note that the exact  two loop expression (\ref{kl2}) is  the coupling
of the 't Hooft scheme).
Numerical results are summarized in the Table 1.
The difference between the solutions in the infrared region
is  less then $1\%$.

There are several reasons to believe that the application of the Lambert
function may  be useful from practical
point of view:

1) Of course, the transcendental equation (\ref{pt}) can be solved
in the complex domain using numerical methods \cite{epl, tau}.
However, this implies a preliminary analytical investigation:
It is necessary to determine the analyticity structure of
the implicit function (the relevant branch should be chosen).
The application of the Lambert W function facilitates
this  task.

2) The explicit inverted 2- and 3-loop solutions of RG equation
may also be useful in studies of renormalon properties of gauge theories
\cite{prf}.

Note added. While preparing this manuscript I became aware of
the paper  \cite{ggk} which has some overlap with the present work.
The authors of \cite{ggk} have obtained explicit expressions for the
coupling (at the 2-loop and 3-loop orders) in terms of the Lambert W function.

\bigskip

\section{The $\beta$-function in the IAA}
It is instructive to consider the $\beta$-function.
From  Eqs.~(\ref{eq:eff}), (\ref{nrm}) and (\ref{kl})  we have
\be
\label{betn}
\beta_{an}^{(n)}(\alpha_{s})=-\frac{1}{\pi}\int_{0}^{\infty}
\frac{{\mu}^2}{(\sigma+{\mu}^2)^2}\rho^{(n)}(\sigma)d\sigma=
-\frac{1}{\pi}\int_{0}^{\infty}\frac{\bar\rho^{(n)}(s)}
{y(s+\frac{1}{y})^2}ds,
\ee
where $\bar\rho^{(n)}(s) \equiv \rho^{(n)}(\sigma)$,
$ s=\frac{\sigma}{{\Lambda}^{2}} $, and $y$
denotes the dimensionless ``nonperturbative'' variable
\be
\label{xi}
y=exp(-\phi)=\frac{{{\Lambda}}^2}{{\mu}^2},
\ee
for $\alpha_{s}\rightarrow 0$,
$y \sim\exp(-\frac{4\pi}{\beta_{0}\alpha_{s}})$ .
The spectral representation (\ref{betn}) defines
$\beta(\alpha_{s})\equiv\hat{\beta}(y)$ as an  analytic function in the
cut-$y$ plane. The cut is along the negative $y$-axis.
Another useful representation for the $\beta$-function follows from
Eqs.~(\ref{nrm}) and (\ref{anc2})
\be
\label{betan}
{\beta}^{(n)}_{an}(\alpha_{s})=\frac{\partial \alpha_{s}}{ \partial ln \mu^2}=
{\beta}^{(n)}_{pt}(\bar\alpha^{(n)}(\mu^2))+\bar{\theta}^{(n)}
(\mu^2)=
{\beta}^{(n)}_{pt}(\alpha_{s}-{\theta}^{(n)}(\mu^2))+
\bar{\theta}^{(n)}(\mu^2),
\ee
where ${\beta}^{(n)}_{pt}$ denotes the nth-order
perturbative approximation for the $\beta$-function
$${\beta}^{(n)}_{pt}(\alpha)=
-\sum_{k=0}^{n-1}\frac{\beta_{k}}{(4\pi)^{k+1}}\alpha^{k+2},$$
and $\bar{\theta}^{(n)}({\mu}^2)={\mu}^2\frac{\partial {\theta}^{(n)}
({\mu}^2)}{\partial {\mu}^2}.$
Eq.~(\ref{betan}) can be rewritten as follows
\be
\label{betan1}
{\beta}_{an}({\alpha}_{s})={\beta}_{pt}({\alpha}_{s})+{\beta}_{np}
({\alpha}_{s},y).
\ee
Here, the nonperturbative piece,  ${\beta}_{np}$,   denotes
the generalized  power expansion for $ {\alpha}_{s} $
\be
\label{np1}
{\beta}_{np}^{(n)}({\alpha}_{s},y)=\sum_{k=0}^{n}
B_{k}^{(n)}(y)(\alpha_{s})^{k},
\ee
where
\be
\label{npc}
\begin{array}{lllll}
B_{0}=\bar\theta,&
B_{1}=-\frac{\partial\beta_{pt}(-\theta)}{\partial\theta},&
B_{k}=\frac{(-1)^k}{k!}\frac{{\partial}^{k}{{\beta}_{pt}}
(-\theta)}{\partial\theta^{k}}+\frac{\beta_{k-2}}{(4\pi)^{k+1}} &
$for$&   k\ge 2.
\end{array}
\ee
Here, for convenience, we  use condensed notations suppressing the
superscript (n) and arguments  to the expressions above
($\theta\equiv{\theta}^{(n)}({\mu}^2)$ ets.).
From the spectral representation (\ref{theta})
and Eq.~(\ref{npc}) we see that
the coefficients
$B_{k}(y)$ are  analytical functions of  $y$
in the cut
$y$-plane. The cut is along the  real positive $y$-axis
$y\ge \frac{1}{k_{L}}>0 $ .
In particular, the coefficients are regular functions
in the  neighborhood of $y=0$.
For  $\alpha_{s}\rightarrow 0$, the coefficients vanish exponentially
${B}_{k}(y)\sim y\sim\exp(-\frac{4\pi}{\beta_{0}\alpha_{s}})$.

A detailed study of the one-loop $\beta$-function can be found in
Ref.~\cite{sh}. It was shown that the one-loop $\beta$-function
obeys the symmetry property
$\beta(\alpha_{s})=\beta(\frac{4\pi}{\beta_{0}}-\alpha_{s})$,
and it has the second order zero at  $\alpha_{s}=\frac{4\pi}{\beta_{0}}$.
To second order the symmetry property for the $\beta$-function is not occurred.

Notice that in the RG IAA  the nonperturbative
$exp(-\frac{1}{a_{s}})$-type terms play an essential role
for restoring of analyticity.
The reason is that one starts with a finite order perturbative
approximation to the $\beta$-function.
Another point of view has been advocated in Refs.~\cite{kp}.
It was shown that a perturbative series for the $\beta$-function,
in the specific class of schemes,  can be summed. So that the
resulting  running coupling has causal structure.
The nonperturbative terms are not added, but  the freedom of choosing
a RS for an infinite series was used instead \cite{kp}.
The connection between that approach and IAA is still unclear.
Indeed, they offer two different causal approximations for
the running coupling.

To illustrate we consider the model $\beta$-function introduced in
Ref.~\cite{crw}
\be
\label{crn}
\begin{array}{ccc}
\bar\beta(h)=-\frac{\beta_{0}h^{2}}{1+\beta_{0}h}:&
h=\frac{\alpha_{s}}{4\pi},& \bar\beta(h)=4\pi\beta(\alpha_{s}),
\end{array}
\ee
which   yields a linearly rising potential and asymptotic freedom.
The RG equation with (\ref{crn}) can be solved, the solution is
\be
\label{crl}
\begin{array}{cc}
\bar h(\frac{Q^2}{{\Lambda}^2})=\frac{1}{\beta_{0} W_{0}\left(\frac{Q^2}
{\beta_{0}{\Lambda}^2}\right)}:&
{\Lambda}^2=h{\mu}^2 exp(-\frac{1}{\beta_{0}h}),
\end{array}
\ee
here $ W_{0}(z)$ is principal branch of $ W(z)$ \cite{lamb}.
Quite remarkably,  (\ref{crl}) satisfies the cut-plane analyticity.
Indeed, $W_{0}(z)$ is analytic at $z=0$ $(W_{0}(z)=z-z^2...)$ and
its branch cut is $\{z:  -\infty<z<-1/e\}$. A generalization of this model
has been studied  in Ref.\cite{ar}.

Generally, the running coupling is a gauge- and scheme-dependent
quantity (it is non-physical object in this sense \cite{kp}).
Presumably, more  useful quantities, in the infrared region,
are the scheme-independent
effective charge (the combination of a vertex and propagators) \cite{cx})
or the running couplings of physical schemes
\cite{gr1,kkp, br}.
Most satisfactory approach can be based on the notion of scheme-
and gauge-independent universal charge (the analogous  of the QED
effective charge) \cite{dmw}.
Recently, in Ref.~\cite{wt} the universal one-loop QCD coupling has been
constructed using the pinch technique.

It is clear that IAA does not exhaust all nonperturbative effects
of the invariant coupling  because its origin is PT.
In Refs.~\cite{dmw, gr, aa} more general frameworks are considered.
It is assumed that a nonperturbative universal QCD coupling  exists
which satisfies the DR.

We can   write the nonperturbative universal coupling as a decomposition
$$
\alpha_{eff}(Q^2)=\alpha_{eff}^{IR}(Q^2)+\alpha_{eff}^{UV}(Q^2).
$$
The infrared and ultraviolet pieces of the coupling satisfy the DRs
\be
\begin{array}{cc}
\alpha_{eff}^{IR}(Q^2)=\frac{1}{\pi}\int\limits_{0}^{K^2(\alpha_{s},\mu)}
\frac{\rho^{IR}(\sigma)}{\sigma+Q^2-i0}d\sigma,&
\alpha_{eff}^{UV}(Q^2)=\frac{1}{\pi}\int\limits_{K^2(\alpha_{s},\mu)}^{\infty}
\frac{\rho^{UV}(\sigma)}{\sigma+Q^2-i0}d\sigma.\\
\end{array}
\ee
The momentum $K(\alpha_{s},\mu)$ divides the infrared and ultraviolet
regions. It may be chosen in a RG-invariant fashion.
In Refs.~\cite{oz} it was shown that
this nonperturbative
scale actually exists in QCD if $N_{f}\leq 9$.
In the weak coupling limit the scale vanishes
exponentially,  $K(\alpha_{s},\mu) \sim \Lambda
\sim e^{-\frac{4\pi}{\beta_{0}\alpha_{s}}}$ as $\alpha_{s}\rightarrow 0$.
The APT approach corresponds to the simplest possibility: it is assumed that
$K=0$, hence $\alpha_{eff}^{ir}(Q^2)$ is ignored. In addition,
the weight function $\rho^{UV}(\sigma)$ is approximated by PT.

One can,  in principle, examine the consistency  of IAA
with a truncated set of SDEs.
The authors of Refs.~\cite{aa} have given
reasons to believe that
$\alpha_{eff}^{IR}(Q^2)$ cannot be  zero, and an infrared regular QCD
effective charge   cannot be consistent with the SDE
for the gluon propagator. Ghost-free axial gauge has been chosen in
this work. On the contrary, the authors of Ref.\cite{sha} have
found that the infrared-regular nonperturbative effective charge
is consistent with the truncated SDEs in the Landau gauge.
However, it was shown that the corresponding  gluon
propagator has not causal structure. We see that situation
is not completely clear.

\bigskip

\section{The gluon propagator in IAA}
In general the gluon propagator is not observable. But, this fact does not
imply that it does not contain physics.
One can relate the propagator to gauge invariant quantities,
for example the Wilson loop \cite{wst}, the gluon condensate \cite{aa} or the
gluon distribution function of hadrons \cite{gg}.
Although the propagator is known to be a gauge variant quantity
it contains an important information about the infrared region
\cite{dl,mn,bbz,ar}.

It has been proved that the propagators in QCD obey the DRs \cite{oz}.
In the standard RG improved PT the DRs  are violated  \cite{oz1}.
Therefore it is desirable  to develop   the analytical approach
for calculating of the propagators.
It was pointed out  in Ref.~\cite{sh1} that there is an ambiguity
due to ``noncommutativity'' of ``analyticization'' with some elements of the
RG algorithm.
For this reason, in the case of Green's functions and observables,
several different versions of ``analyticization'' are
possible.
To illustrate we shall consider the gluon propagator in the Landau gauge
\be
\begin{array}{ccc}
D_{\mu\nu}(Q)=(g_{\mu\nu}+\frac{Q_{\mu}Q_{\nu}}{Q^2})D(Q^2):
& D(Q^2)=\frac{d(\frac{Q^2}{{\mu}^2},\alpha_{s})}{Q^2},&d(1,g^2)=1.
\end{array}
\label{eq:gl}
\ee
We have assumed that quarks are massless and normalized the propagator at the
Euclidean point $Q^2={\mu}^2$.
In the Landau gauge, invariance under the RG  leads
\be
\label{gs}
d\left(\frac{Q^2}{\mu^2}, \alpha_{s}\right)=\exp\left(\int_{\alpha_{s}}^
{\bar \alpha_{s}
(Q^2)}\frac{\gamma_{v}(x)}
{\beta(x)}dx\right).
\ee
The formal power series for the anomalous dimension function
$\gamma_{v}$ (in this  gauge) is given by
\be
\begin{array}{cc}
\label{gv}
{\gamma}_{v}(\alpha_{s})=-(\frac{{\gamma_{0}}}{4\pi}\alpha_{s}+
\frac{\gamma_{1}}{16{\pi}^2}{\alpha_{s}}^2
+\ldots), & {\gamma}_{0}=\frac{1}{2}(13-\frac{4}{3}N_{f}).
\end{array}
\ee
A detailed investigation of the gluon propagator in the framework of
analyticity and asymptotic freedom has been undertaken in
Refs.\cite{oz}. It has been shown that the
propagator amplitude satisfies an unsubtracted DR
\be
\label{klg}
D(Q^2, \alpha_{s},\mu)=
\frac{1}{\pi}\int\limits_0^{\infty}
\frac{\rho_{v}(\sigma,\alpha_{s},\mu)}
{\sigma+Q^2-i0}d\sigma.
\ee
For a limited number of flavors $N_{f}\leq 9 $,
in the Landau gauge,
the weight function $\rho_{v}$ obeys the superconvergence relation
\be
\label{sr}
\int\limits_0^{\infty}\rho_{v}(\sigma,\alpha_{s},\mu) d\sigma= 0.
\ee
This  relation is considered as a sufficient  condition
for color  confinement in  the approach
to a confinement based on the BRST algebra and the RG
\cite{o1,nis}.
The consequence of  (\ref{sr}) is that there is a renormalization
invariant point $K^{2}(\alpha_{s},\mu)$ such that $\rho_{v}$ is a negative
measure for $\sigma\geq K^2$ \cite{oz}.

Let us consider few different scenarios of restoring  analyticity
for the gluon propagator:

(a) The  Analytic Perturbation Theory (APT) possibility.
This specific recipe
for ``analyticization'' of an observable has been introduced and elaborated
in Refs.\cite{tau,bjo, epl}.
Here, instead of the power perturbation series, an amplitude is
presented in a form of an asymptotic expansion of a more general form,
the expansion over an asymptotic set of functions $[a^{n}(x)]_{an}$,
the ``n-th power of $a(x)$ analyticized as a whole'' \cite{sh1}.
In the APT approach, the drastic reduction of loop and RS
sensitivity for several observables has been found.

In this case the starting point is formula (\ref{gs}).
In the one-loop approximation  (\ref{gs}) yields
\be
\label{gpt}
D^{(1)}(\frac{Q^2}{{\mu}^2},\alpha_{s})=
\frac{1}{Q^2}\left(\frac{\bar\alpha^{(1)}(Q^2)}{\alpha_{s}}\right)^
{\frac{\gamma_{0}}{\beta_{0}}}=
\frac{1}{Q^2}\left(\frac{4\pi}{\alpha_{s}\beta_{0}\ln\left(\frac{Q^2}
{\Lambda^{2}}\right)}\right)^{\frac{\gamma_{0}}{\beta_{0}}}.
\ee
Since the ratio $\frac{\gamma_{0}}{\beta_{0}}$ is fractional number,
expression (\ref{gpt}) has the ghost cut $0<Q^2<\Lambda^2$.
The discontinuity $\rho_{v}^{(1)}$ of $ D^{(1)}$ along
the negative $Q^2$ axis
can be written as \cite{oz1}
\be
\label{ro1}
\begin{array}{cc}
\rho_{v}^{(1)}(\sigma)=\displaystyle\frac{\bar\rho_{v}^{(1)}(s)}{\Lambda^2};&
\bar\rho_{v}^{(1)}(s)=-c_{v}(\alpha_{s})
\displaystyle\frac{1}{s}
(R(s))^{\frac{\gamma_{0}}{\beta_{0}}}
\sin\left[\frac{\gamma_{0}}{\beta_{0}}\Phi(s)\right],
\end{array}
\ee
where $s=\frac{\sigma}{\Lambda^{2}}$,
$c_{v}=\left(\frac{4\pi}{\beta_{0}\alpha_{s}}\right)^
{\frac{\gamma_{0}}{\beta_{0}}}$,
$R(s)=(\ln^{2} s+\pi^{2})^{-\frac{1}{2}}$ and
$$
\Phi(s)=\left\{ \begin{array}{ll}\arcsin(\pi R(s))& \mbox{if $s>1$} \\
\pi-\arcsin(\pi R(s))& \mbox{if $0<s<1$}.
\end{array}
\right.
$$
To construct the corresponding analytic expression one has to substitute
(\ref{ro1}) in formula  (\ref{klg}).
Note that the weight function (\ref{ro1}) has a nonintegrable singularity
at $\sigma=0$. So that the  spectral representation (\ref{klg})
with  the weight function (\ref{ro1}) diverges.
Nevertheless, in  sense
of the distribution theory this problem may be solved
\footnote{The infrared behavior of spectral representations has been studied
in Refs.~\cite{oz}.}.
Using the method of Refs.~\cite{oz} we find the following
subtracted DR for the ``analyticized'' amplitude
\be
\label{subtr}
D_{an1}^{(1)}(Q^2)=\frac{c_{1}}{Q^2}-\frac{1}{Q^2\pi}
\int\limits_0^{1} \frac{s \bar\rho_{v}^{(1)}(s) ds}
{s+\frac{Q^2}{{\Lambda}^2}}
+\frac{1}{{\Lambda}^2\pi}\int\limits_1^{\infty}
\frac{\bar\rho_{v}^{(1)}(s) ds}
{s+\frac{Q^2}{{\Lambda}^2}},
\ee
where
\be
\label{coef}
\begin{array}{ccc}
c_{1}=\frac{1}{\pi}\int\limits_0^{1} \bar \rho_{v.reg}^{(1)}(s) ds; &
\bar\rho^{(1)}_{v.reg}=\bar\rho_{v}^{(1)}-
\bar\rho_{v.sing}^{(1)};&
\bar\rho_{v.sing}^{(1)}(s)=
-\displaystyle\frac{c_{v}(\alpha_{s})}{s|\ln s|^p}\sin(p\pi),
\end{array}
\ee
with $p=\frac{\gamma_{0}}{\beta_{0}}=\frac{39-4N_{f}}{2(33-2N_{f})}$
($ 0<p<1$ if $ N_{f}\leq 9$), and $\bar\rho_{v}^{(1)}(s)$ is given by (\ref{ro1}).
Instead of the superconvergence relation  (\ref{sr})
now we have the following relation \cite{oz1}
\be
\label{nwsr}
 \int\limits_0^{1} \bar \rho_{v.reg}^{(1)}(s) ds
    + \int\limits_1^{\infty} \bar \rho_{v}^{(1)}(s)ds=0.
\ee

However, there is another way to handle the above considered
infrared problem. The point is, that for $N_{f}\leq 9$
one can write an unsubtracted DR for the dimensionless structure
$d$, $d=Q^2D(Q^2)$ \cite {oz}.
This allows us to construct the analytic expression for $d$.
The corresponding ``analyticized'' amplitude $D_{an}$ is then determined by
\be
\label{unsubt}
D_{an1}^{(1)}(Q^2)=\frac{d_{an}^{(1)}(\tilde Q^2)}{Q^2}=
\frac{1}{Q^2\pi}\int\limits_0^{\infty}
\frac{ \bar \rho_{v1}^{(1)}(s) ds}{s+{\tilde Q}^2}
\ee
where ${\tilde Q}^2=\frac{Q^2}{\Lambda^2}$,
$\bar\rho_{v1}^{(1)}(s)=-s\bar\rho_{v}^{(1)}(s)>0$ with
$\bar\rho_{v}^{(1)}(s)$  given by  Eq.(\ref{ro1}).
One can verify that expression (\ref{unsubt}) has the correct
``abelian limit''. Indeed, we have
$$
\lim_{\gamma_{0}\rightarrow\beta_{0}}a_{s}d_{an}^{(1)}({\tilde Q}^2)
= \frac{1}{\pi}\int\limits_0^{\infty}
\frac{ds}{s+{\tilde Q}^2}\left(\frac{\pi}{\ln^2 s+{\pi}^2}\right)=
a_{an}^{(1)}({\tilde Q}^2)\equiv\left(\frac{1}{\ln {\tilde Q}^2}+
\frac{1}{1-{\tilde Q}^2}\right).
$$
Using the relation (\ref{nwsr}) we verify that the analytic
expression (\ref{unsubt}) follows from the representation
(\ref{subtr}).

Note that (\ref{unsubt}) does not possess a pole at $Q^2=0$ and it
has a more strong singularity then the corresponding amplitude
for the free field. By direct calculation we find
$$
\begin{array}{ccc}
D_{an1}^{(1)}(Q^2)\sim\displaystyle\frac {c_{v}}{\pi(1-p)}
(|\ln(\frac{Q^2}{{\Lambda}^2})|)^{1-p}\displaystyle\frac{1}{Q^2} &\mbox{for}
 &Q^2\rightarrow 0.
\end{array}
$$
More serious difficulty is that there is not proper
connection between the
analytic propagator (\ref{subtr}) and the analytic running
coupling as it follows from basic RG relations \cite{sh1}.
Indeed, the product of a vertex and appropriate
powers of propagators should form an invariant charge.

(b) One can substitute the explicit expressions $a_{an}(Q^2)$
in formula (\ref{gs}).
The simple possibility is to  use the perturbative expression (\ref{gv}) for
$\gamma_{v}$. This choice has been proposed in Ref.~\cite{nst}.
Then in the one-loop approximation formula (\ref{gs})
gives the causal expression
\be
\label{nst}
D^{(1)}_{an2}(Q^{2})=\frac{c_{v2}}{Q^{2}}
\left[\frac{{\tilde Q}^{2}-1}{{\tilde Q}^{2}\ln {\tilde Q}^{2}}\right]^
{\frac{\gamma_{0}}{\beta_{0}}},
\ee
where $\tilde Q=\frac{Q}{\Lambda}$, and $c_{v2}$ is the normalization constant.
As in the case (a), we see that (\ref{nst}) does not satisfy the
unsubtracted DR (\ref{klg}). The lack of this solution is that
it does not reproduce  the ``abelian limit'' for
$\gamma_{0}\rightarrow\beta_{0}$.

(c) We may introduce RG improved power asymptotic expansion
differing from the usual one
by substitution $a_{pert.}(\frac{Q^2}{{\Lambda}^2})\Rightarrow
a_{an.}(\frac{Q^2}{{\Lambda}^2})$ only.
With this choice, in the one-loop order, we get
\be
\label{1lgs}
 D^{(1)}_{an3}(Q^2)=\frac{1}{Q^2}\left(\frac{\bar\alpha_{an}^{(1)}
 (Q^2)}
 {\alpha_{s}}\right)^{\frac{\gamma_{0}}{\beta_{0}}}=
 \frac{1}{Q^2}c_{v}\left(\frac{1}{\ln{\tilde Q}^2}+
 \frac{1}{1-{\tilde Q}^2} \right)^{\frac{\gamma_{0}}{\beta_{0}}},
   \ee
where    $c_{v}=(\frac{4\pi}{\beta_{0}\alpha_{s}})^
{\frac{\gamma_{0}}{\beta_{0}}}$ and ${\tilde Q}^2=\frac{Q^2}{\Lambda^2}$.
It is easy to convince that (\ref{1lgs})  has correct
analytical properties.
Indeed, $\bar\alpha_{an}^{(1)}(Q^2) $  does not
vanish  in the finite part of the complex $Q^2$-plane.
Moreover, (\ref{1lgs}) satisfies the unsubtracted DR
(see Refs.~\cite{my,my1}).
For the RG invariant scale $K(g,\mu^2)$,
the solution (\ref{1lgs}) yields trivial value $K=0$.
In a similar way, one can also obtain causal
approximations to $D(Q^2)$ at higher orders.

(d) Consider the one-loop gluon self energy
$\Pi^{(1)}$,
\be
\Pi^{(1)}({\bar Q}^{2},a_{s})=\frac{\gamma_{0}}{\beta_{0}}a_{s}
\ln {\bar Q}^2,
\ee
here $\gamma_{0}$ is given by (\ref{gv}),
$a_{s}=\frac{\beta_{0}}{4\pi}\alpha_{s}$ and ${\bar Q}^2=
\frac {Q^{2}}{\mu^{2}}$.
We can introduce the Dyson series in $\Pi^{(1)}$ for the propagator.
Summing this series we get
\be
\label{ds}
d_{Ds}({\bar Q}^{2},a_{s})=
\frac{1}{1+\Pi^{(1)}({\bar Q}^{2},a_{s})}=
-\kappa\ln y \frac{1}{\ln \frac{{\bar Q}^{2}}{y^{\kappa}}},
\ee
here $\kappa=\frac{\beta_{0}}{\gamma_{0}}$ and instead of $a_{s}$ we have
introduced the variable $y$, $y=e^{-\frac{1}{a_{s}}}$.
The approximation  (\ref{ds}) does not satisfy RG invariance
and has the ghost pole.
We may define the corresponding ``analyticized'' amplitude \cite{red}
\be
\label{dsan}
d_{Ds.an}({\bar Q}^{2},y)=
\frac{1}{1+{\Pi}^{(1)}_{mod}({\bar Q}^{2},y)}=
C_{Ds}(y)\left(\frac{1}{\ln ({\bar Q}^{2}y^{-\kappa})}
-\frac{y^{\kappa}}{{\bar Q}^2-y^{\kappa}}\right),
\ee
in order to preserve the normalization (see (\ref{eq:gl})) we have
introduced in (\ref{dsan})
the necessary  factor $c_{Ds}(y)$.
The effect of this procedure is that the gluon self-energy receives
a pure nonperturbative contribution.
The modified gluon self-energy function is  then given by
\be
\label{pmod}
\Pi^{(1)}_{mod}({\bar Q}^2, y)=\Pi^{(1)}({\bar Q}^2, -{(\ln y)}^{-1})
+\Pi^{(1)}_{nonpert.}({\bar Q}^2, y),
\ee
here
\be
\Pi^{(1)}_{nonpert.}({\bar Q}^2, y)=
(1+\Pi^{(1)}({\bar Q}^2))
\left(-1+\frac{(\kappa\ln y)^{-1}+G(1)}
{(\kappa\ln y)^{-1}+G({\bar Q}^2)(1+\Pi^{(1)}({\bar Q}^2))}\right)
\ee
with $\Pi^{(1)}({\bar Q}^2)\equiv\Pi^{(1)}({\bar Q}^2, -{(\ln y)}^{-1}) $
and
$G({\bar Q}^2)=y^{\kappa}({\bar Q}^2-y^{\kappa})^{-1}$.
We see that the expression (\ref{pmod}) is free from the ghost singularity.
For  $N_{f}\leq 9$ $(\kappa>0)$,
in the weak-coupling limit,
the nonperturbative part of the self energy vanishes exponentially:
$\Pi^{(1)}_{nonpert.}({\bar Q}^2, y)\sim O(e^{-\frac{\kappa}{a_{s}}})
\rightarrow 0$.
The modified one loop amplitude is then given by
\be
\label{mod}
d^{(1)}_{mod}({\bar Q}^2,y)=1- \Pi^{(1)}_{mod}({\bar Q}^2, y).
\ee
Now, the perturbative definition of $y$,
$ y=e^{-\frac{1}{a_{s}}} $, is not valid. Instead, we shall accept
the modified relation
$y=\frac{\Lambda^{2}}{\mu^{2}}=e^{-\phi(a_{s})}$ (see Refs.\cite{bog,sh}).
Thus, to first order, $ \phi(a_{s})$ is determined by \cite{sh}
$$
\label{scl}
\frac{1}{\phi(a_{s})}+\frac{1}{1-e^{\phi(a_{s})}}=a_{s}.
$$
We can calculate the modified
anomalous dimension for the gluon field
\be
\gamma_{mod}^{(1)}(a_{s})=\bar{\gamma}_{mod}(\phi)=
\lim_{{\bar Q}^2\rightarrow 1} {\bar Q}^2 \frac{\partial}{\partial {\bar Q}^2}
\ln d^{(1)}_{mod}({\bar Q}^2,y)=
\lim_{{\bar Q}^2\rightarrow 1} {\bar Q}^2 \frac{\partial}{\partial {\bar Q}^2}
\ln d_{Ds.an}({\bar Q}^2,y),
\ee
using (\ref{dsan}) we find
\be
\label{adim}
\bar{\gamma}_{mod}(\phi)=\frac{1}{\kappa}\frac{d}{d\phi}
\ln \left(\frac{1}{\kappa\phi}-\frac{1}{e^{\kappa\phi}-1}\right).
\ee
Inserting expression (\ref{adim}) in Eq. (\ref{gs})  we obtain
the corresponding RG improved expression for the propagator
amplitude
\be
\label{sl3}
D_{RG}(Q^2)=
\frac{d_{RG}\left(\frac{Q^2}{{\Lambda}^2}, y\right)}{Q^2}=
\frac{C_{RG}(y)}{Q^2}\left(\frac{1}{\kappa\ln\left(\frac{Q^2}
{{{\Lambda}}^2}\right)}
-\frac{1}{\left(\frac{Q^2}{{{\Lambda}}^2}\right)^{\kappa}-1}
\right)^{\frac{1}{\kappa}},
\ee
here, the factor $C_{RG}(y)$ is determined by the normalization
condition  (\ref{eq:gl}).
We see that, in the weak-coupling limit  $a_{s}\rightarrow 0$
(for $N_{f}\leq 9$) expression (\ref{sl3}) reproduces the standard
RG improved solution (\ref{gpt}).
Furthermore, it has the correct  ``abelian limit''
$a_{s}d_{RG}(\frac{Q^2}{\Lambda^2},y)\rightarrow a_{an}(\frac{Q^2}{\Lambda^2})$
as $\frac{\beta_{0}}{\gamma_{0}}\rightarrow 1$.

For $\frac{\gamma_{0}}{\beta_{0}} \ge 0.5$,
the function (\ref{sl3}) satisfies the
K\"{a}llen-Lehmann analyticity.
The corresponding flavor condition is $N_{f}\le3$.
For $N_{f}>3$, the unphysical singularities appear in
the first Riemann sheet.
This limitation for $N_{f}$ turns out to be natural.
Indeed, the singularities in (\ref{sl3}) occurred at
$|\frac{k^2}{{{\Lambda}}^2}|=1$ where
the number of active quarks are just three.
On the other hand, the flavor condition $ N_f\le3 $
seems to be plausible for massless QCD.
For this reason we cannot
reject the solution (\ref{sl3}) using arguments of analyticity
\footnote{The author owes to D.V. Shirkov for drawing his attention
to this peculiarity of the solution (\ref{sl3}).}.

It is interesting to compare the  analytic solutions  (\ref{subtr}),
 (\ref{1lgs}) and  (\ref{sl3}), which have correct ``abelian limit''.
 Only (\ref{1lgs}) and  (\ref{sl3}) satisfy an unsubtracted DR.
The convenient  criterion has been formulated in Refs.\cite{aa}.
This is the principle of minimality for the nonperturbative contributions
in perturbative (ultraviolet) region.
According to this principle one can easily verify that the   solution
(\ref{sl3}) is preferable. Indeed,
 it predicts more rapid decrease of the nonperturbative contributions
in the ultraviolet region  then the solutions (\ref{subtr}), (\ref{nst}) and
(\ref{1lgs}).
However, we cannot accept a final decision for selecting the solutions.
Indeed, the above mention principle is relevant to the full gluon
propagator,  which may include   a pure nonperturbative
contributions. In the IAA these contributions are invisible.

\section{Conclusion}
The  RG equation for the  QCD  two-loop invariant  coupling
has been solved explicitly.
We have expressed the solution  in terms of the Lambert W function.
This allows us to understand more clearly  the analytical structure
of the solution in the complex $Q^2$ plane.
The corresponding analytic coupling has been reconstructed via the
dispersion relation.
It has been demonstrated that the ``analyticized'' iterative solution
(\ref{kl1})   is
numerically close to the ``analyticized'' exact one (\ref{kl2}).

We have expressed the invariant (running) coupling of the special RS
via the Lambert W function. The corresponding analytic invariant
coupling is constructed (see (\ref{kl3})).
We have shown that the analytic (running) couplings of the  two
considered schemes are    numerically close in the IR region.

The  structure of the   $\beta$-function
has been analyzed in IAA.
We have  solved the  RG equation,  with nonperturbative model
$\beta$-function,
giving explicit expression for the invariant coupling as a function
of the scale  also in terms of the Lambert W function (see (\ref{crl})).
We have found that the solution is automatically causal.

The one-loop gluon propagator amplitude of massless QCD is considered
in the Landau gauge.
The RG and analyticity constraints alone are not sufficient to
uniquely determine the analytic solution for the gluon propagator
starting from PT. Therefore,
several  versions of ``analyticization'' of the gluon propagator are considered.
Properties of the obtained analytical solutions for the propagator are
discussed.

Finally,
we remark that the gluon propagator is central object in the
framework based on the SDEs  \cite{mn,bbz,ar}. Here,
the analytic perturbative solutions can be used to derive more
complete (nonperturbative) approximants to the gluon propagator.
For related ideas and applications, see \cite{aa}.

{\bf Acknowledgments}\\
The author wish to thank D.V. Shirkov for  kind hospitality
in Dubna,  for very helpful discussions and for critical reading of
the manuscript.
It is a pleasure to acknowledge
B.A.~ Arbuzov, R.~ Bantsuri, M.A.~ Eliashvili,  G.P.~ Jorjadze,
A.L.~ Kataev, D.I.~ Kazakov,  A.A.~ Khelashvili, A.M.~ Khvedelidze,
A.N.~ Kvinikhidze,  G.V.~ Lavrelashvili, V.A.~ Matveev, A.V.~ Nesterenko, A.A.~ Pivovarov,
I.L.~ Solovtsov,  O.P.~ Solovtsova, A.N.~ Tavkhelidze
for valuable discussions on this topic.\\

\end{document}